\documentclass[10pt, a4paper]{article}

\usepackage{graphics} 
\usepackage{mathptmx} 
\usepackage{times} 
\usepackage{amsmath} 
\usepackage{amssymb}  
\usepackage{color}
\usepackage{hyperref}
\usepackage[dvips]{graphicx}
\begin{document}

\begin{flushleft}
{\large
\textbf{Synergetic and redundant information flow detected by unnormalized Granger causality: application to resting state fMRI}
}
\\
\vspace{5pt}
{\small
Sebastiano Stramaglia$^{1,2}$, 
Leonardo Angelini$^{1}$,
Guorong Wu$^{3,4}$,
Jesus Cort{\'e}s$^{5}$,
Luca Faes$^{6}$,
Daniele Marinazzo$^{4,\ast}$
}
\\
\vspace{5pt}
{\scriptsize
\bf{1} Dipartimento Interateneo di Fisica, University of Bari, and INFN Sezione di Bari, Italy
\\
\bf{2} Basque Center for Applied Mathematics, Bilbao, Spain
\\
\bf{3} BIOtech, Dept. of Industrial Engineering, University of Trento, and IRCS-PAT FBK, Trento, Italy
\\
\bf{4}  Computational Neuroimaging Lab, Biocruces Health Research Institute, Cruces University Hospital, Barakaldo,
Spain and Ikerbasque, The Basque Foundation for Science, Bilbao, Spain
\\
\bf{5} Faculty of Psychology and Educational Sciences, Department of Data Analysis, Ghent University, Belgium
\\
\bf{5} BIOtech, Dept. of Industrial Engineering, University of Trento, and IRCS-PAT FBK, Trento, Italy
\\
$\ast$ E-mail: daniele.marinazzo@ugent.be
}
\end{flushleft}

\begin{abstract}
\textit{Objectives:} We develop a framework for the
analysis of synergy and redundancy in the pattern of information
flow between subsystems of a complex network. \textit{Methods:} The
presence of  redundancy and/or synergy in multivariate time series
data  renders difficult to estimate the neat flow of information
from each driver variable to a given target. We show that adopting
an unnormalized definition of Granger causality one may put in
evidence redundant multiplets of variables influencing the target by
maximizing the total Granger causality to a given target, over all
the possible partitions of the set of driving variables.
Consequently we introduce a pairwise index of synergy which is zero
when two independent sources additively influence the future state
of the system, differently from previous definitions of synergy.
\textit{Results:} We report the application of the proposed approach to
resting state fMRI data from the Human Connectome Project, showing
that redundant pairs of regions arise mainly due to space contiguity
and interhemispheric symmetry, whilst synergy occurs mainly between
non-homologous pairs of regions in opposite hemispheres.
\textit{Conclusions:} Redundancy and synergy, in healthy resting brains,
display characteristic patterns, revealed by the proposed approach.
\textit{Significance:} The pairwise synergy index, here introduced, maps
the informational character of the system at hand into a weighted
complex network: the same approach can be applied to other complex
systems whose normal state corresponds to a balance between
redundant and synergetic circuits.
\end{abstract}


%

\section{Introduction}

The inference of dynamical networks from time series data is
related to the estimation of the  information flow between variables
\cite{pereda2005}. Granger causality (GC) \cite{granger} has emerged
as a major tool to address this issue. This approach is based on
prediction: if the prediction error of the first time series is
reduced by including measurements from the second one in the linear
regression model, then the second time series is said to have a
Granger causal influence on the first one. An important application
of this notion is neuroscience, see, e.g.,
\cite{baccala,chavez,plompe,porta}.

Interactions in dynamical networks analysis have also been studied with state space models \cite{barnett2015}, or using Kalman filters
\cite{havlicek} to deal with multivariate system identification.
Other approaches are rooted in information theory
\cite{schreiber,lehnertz} and have been applied to the analysis of
complex spatio-temporal systems, e.g. Earth's climate \cite{runge}. Whilst these
methods are effective in the detection and quantification of directed interactions,
they are not suitable to elucidate the informational character of groups
of variables, i.e. whether a group of variables provides information about
a given target in a synergetic, redundant, or independent way. Synergetic here means that they convey more information in their
joint responses than the sum of their individual informational contributions;
redundant means that they jointly convey less than that. Redundancy and
synergy are intuitive yet elusive concepts, which have been
investigated in different fields, including pure information theory
\cite{griffith,harder,Lizier,wibral,faes}. The role of synergy and
redundancy has also been investigated in epilepsy \cite{asier} and disorders of consciousness \cite{massimini}. Those
studies suggest that complex systems in normal conditions tend to
display a proper balance between redundant and synergetic
circuits, and pathologies to disrupt it.

Coming back to Granger causality, it is worth recalling that
its pairwise version consists in assessing influence between each pair
of variables, independently of the rest of the system. It is well
known that the pairwise analysis cannot disambiguate direct and
indirect interactions among variables, neither can detect synergy. We remark that
the synergetic effects that we consider here, related to the
analysis of dynamical influences in multivariate time series, are
similar to those encountered in sociological and psychological
modeling, where {\it suppressors} is the name given to variables
that increase the predictive validity of another variable after  its
inclusion into a linear regression equation \cite{conger}. Some
information-based approaches addressing the issue of collective
influence are \cite{Chicharro,Lizier}. The most
straightforward extension of pairwise Granger causality, the
conditioning approach,
removes indirect influences by evaluating to which extent the
predictive power of the driver on the target decreases when the
conditioning variable is removed. Sometimes though a full conditioning can
encounter conceptual limitations, on top of the practical and
computational ones: in the presence of redundant variables the
application of the standard analysis leads to underestimation of
influences \cite{noipre}.  As a convenient alternative to
this suboptimal solution, a partially conditioned approach,
consisting in conditioning on a small number of variables, chosen as
the most informative ones for the driver node, has been proposed
\cite{partialold}.

The purpose of the present work is to show that,
taking into account the above mentioned problems encountered by the
standard definition of Granger causality, an {\it unnormalized}
Granger causality index is better suited for the analysis of systems
of many variables, in presence of synergy and redundancy, to provide
information for the future state of the system. The novel
approach presented here provides a further description of complex
systems where the informational character of multiplets of variables
is highlighted. The proposed method in principle may be applied to
any system characterized by complex regulatory mechanisms.

The paper is organized as follows. The next section is devoted to a
brief account on Granger causality, with particular attention to the
problems which arise due to redundancy and synergy. In Section III
we describe our definition of {\it unnormalized} Granger causality,
providing some examples to show that the proposed metrics
disentangles independent sources of information. In Section IV we
introduce the Pairwise Synergy Index, a weighted network associated
to the informational character of a set of variables, which allows
to use methods of complex networks to analyze the informational
pattern of large data sets; we will show the application of the
proposed approach to resting state fMRI data. Some conclusions will
be drawn in section V.

\section{Granger causality}

Granger causality is a powerful and widespread data-driven  approach
to determine whether and how two time series exert direct dynamical
influences on each other \cite{hla}. A convenient nonlinear
generalization of GC has been implemented in \cite{noipre2},
exploiting the kernel trick, which makes computation of dot products
in high-dimensional feature spaces possible using simple functions
(kernels) defined on pairs of input patterns. This trick allows the
formulation of nonlinear variants of any algorithm that can be cast
in terms of dot products, for example Support Vector Machines
\cite{vapnik}. Hence in \cite{noipre2} the idea is still to perform
linear GC, but in a space defined by the nonlinear features of the
data. This projection is conveniently and implicitly performed
through kernel functions \cite{shawe} and a statistical procedure is
used to avoid overfitting.

Quantitatively, let us consider $n$ time series $\{x_\alpha
(t)\}_{\alpha =1,\ldots,n}$; the lagged state vectors are denoted
$$X_\alpha (t)= \left(x_\alpha (t-m),\ldots,x_\alpha (t-1)\right),$$
$m$ being the order of the model (window length). Let $\epsilon
\left(x_\alpha |{\bf X}\right)$ be the mean squared error prediction
of $x_\alpha$ on the basis of all the vectors ${\bf
X}=\{X_\beta\}_{\beta =1}^n$ (corresponding to the kernel approach
described in \cite{noiprl}). The fully conditioned GC index
$\delta_{mv} (\beta \to \alpha )$ is defined as follows: consider
the problem of predicting $x_\alpha$ on the basis of all the
variables but $X_\beta$ and the problem of predicting $x_\alpha$
using all the variables, then the GC is the logarithm variation of
the prediction  error in the two conditions, i.e.
\begin{equation}\label{mv}
\delta_{mv} (\beta \to \alpha )= \log{\epsilon \left(x_\alpha |{\bf
X}\setminus X_\beta\right)\over \epsilon \left(x_\alpha |{\bf
X}\right)}.
\end{equation}
In \cite{ancona} it has been shown that not all the kernels are
suitable to estimate GC. Two important classes of kernels which can
be used to construct nonlinear GC measures are the {\it
inhomogeneous polynomial kernel} (whose features are all the
monomials in the input variables up to the $p$-th degree; $p=1$
corresponds to linear Granger causality) and the {\it Gaussian
kernel}. We also remark that the complexity of the regression model
can be controlled as explained in \cite{noiprl}, hence the causality
values may be assumed to be  not affected by overfitting.

The pairwise Granger causality is given by:
\begin{equation}\label{bv}
\delta_{bv} (\beta \to \alpha )= \log{\epsilon \left(x_\alpha
|X_\alpha \right)\over \epsilon \left(x_\alpha |X_\alpha ,
X_\beta\right)}.
\end{equation}

The following examples show that conditioned GC tends to be reduced
in presence of redundancy and increased in presence of synergy, the
latter occurrence being a problem for pairwise GC, see
\cite{wibral,noinjp}.

\subsection{Redundancy due to a hidden source}
\label{rrr} 
\begin{figure}
\centering
\includegraphics[width=8cm]{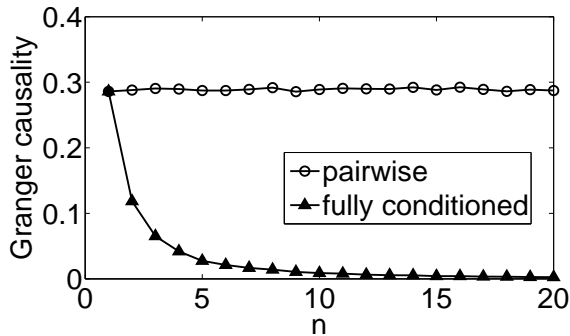}
      \caption{Fully conditioned and pairwise Granger causality are plotted versus the number
of variables for the example regarding redundancy. The causality
displayed is from one of variables $x$ to the target $w$. Results
are averaged over 100 runs of 1000 time points. In both cases the
regression model is linear.}
      \label{fig1}
\end{figure}

We show here how   redundancy constitutes a problem for
fully conditioned GC. Let $h(t)$ be a zero mean and unit variance
hidden Gaussian variable, influencing $n$ variables $x_i (t)=h(t-1)+
s \eta_i (t)$, and let $w(t)=h(t-2) +s \eta_0 (t)$ be another
variable which is influenced by $h$ but with a larger delay. The
$\{\eta\}$ variables are unit variance Gaussian noise and s controls
the noise level. In figure (\ref{fig1}) we depict both the linear
fully connected and pairwise GC from one of the x's to w (note that
h is not used in the regression model). As $n$ increases, the fully
conditioned GC vanishes as a consequence of redundancy, whilst the
GC relation is found for any $n$ in the pairwise analysis.

\subsection{Synergetic contributions}
\begin{figure}
\centering
\includegraphics[width=8cm]{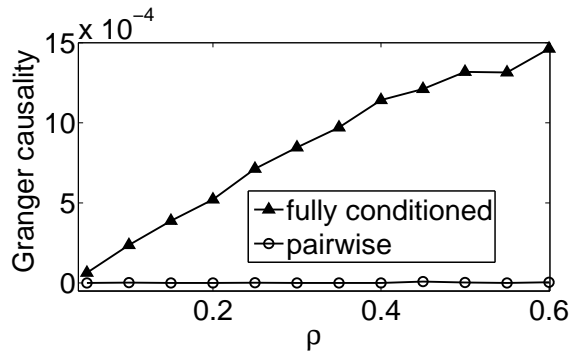}
      \caption{For the example dealing with synergy, CGC and PWGC are plotted
versus the coupling $\rho$, for the causality $x_3 \to x_4$. Results
are averaged over 100 runs of 1000 time points. The regression model
here corresponds to the polynomial kernel of order two in the
approach\cite{noiprl}, i.e. the features are all the monomials in
the input variables up to the second order. }
      \label{fig2}
\end{figure}

Let us consider three unit variance iid Gaussian noise terms $x_1$,
$x_2$ and $x_3$. Let $$x_4(t)=0.1 (x_1(t-1) + x_2(t-1)+\eta(t))+\rho
x_2(t-1)x_3(t-1).$$ Considering the influence $3\to 4$, the fully
conditioned GC reveals that $3$ is influencing $4$, whilst pairwise
GC fails to detect this causal relationship, see figure
(\ref{fig2}), where we use the method described in \cite{noipre}
with the inhomogeneous polynomial kernel of degree two; $x_2$ is a
suppressor variable for $x_3$ w.r.t. the influence on $x_4$. This
example shows that pairwise analysis fails to detect synergetic
contributions. We also remark that use of nonlinear GC is mandatory
in this case to evidence the synergy between $x_2$ and $x_3$.

\section{Methods}
\subsection{Unnormalized Granger Causality}
\label{un} As the two examples at the end of last
section show,  the presence of relevant synergy and/or redundancy in
the data influences the output of standard Granger causal analysis, thus calling
for effective methods to deal with those cases.

Firstly, we remark that interaction information is a classical
measure of the amount of information (redundancy or synergy) bound
up in a set of three variables \cite{mcgill,sch}. In
\cite{preExpansion} a generalization of the interaction information,
to the case of lagged interactions, has been proposed together with
an expansion that allows to extend the definition to any number of
variables. As the sign of the interaction information  corresponds
to synergy or redundancy, this interpretation implies that synergy
and redundancy are taken to be mutually exclusive qualities of the
interactions between variables \cite{beggs}. Other approaches
instead regard synergy and redundancy as separate entities, for
example in \cite{beer} a {\it partial information decomposition}
(PID) was proposed: the information that two 'source' variables Y
and Z hold about a third 'target' variable X can be decomposed into
four parts: (i)the unique information that only Y (out of Y and Z)
holds about X; (ii) the unique information that only Z holds about
X; (iii) the redundant information that both Y and Z hold about X;
and (iv) the synergistic information about X that only arises from
knowing both Y and Z. These quantities have been evaluated
analytically for Gaussian systems \cite{barrett}, and lead to some
undesirable results, e.g., redundancy reduces to the minimum
information provided by either source variable, and hence is
independent of correlation between sources. As suggested in
\cite{barrett}, this occurrence may be related to the fact that
Shannon information between continuous random variables is more
precisely based on differential, and the limit to the continuum is
not straightforward \cite{cufaro}. Therefore in the case of
continuous variables we propose here to describe the informational
character of a subset of variables in terms of the reduction of
variance of residuals of the target due to inclusion of driver
variables, along the lines described in \cite{noipre}. The
informational character of each multiplet will be associated to a
single number, which may be seen as the difference of redundancy and
synergy in every formalism where these two notions are separately
defined (see the discussion in \cite{beggs}).

First of all we note that the straightforward generalization of
Granger causality for driving sets of variables is
\begin{equation}\label{sets}
\delta_{\bf X} (B \to \alpha)=\log{{\epsilon \left(x_\alpha |{\bf
X}\setminus B\right)\over  \epsilon \left(x_\alpha |{\bf
X}\right)}},
\end{equation}
where $B$ are is a subset of variables, $x_\alpha$ is the target
variable and ${\bf X}\setminus B$ means the set of all variables
except for those $X_\beta$ with $\beta\in B$. Note that we have
isolated the variable $X_\alpha$, i.e. the present state of the
target. The subscript ${\bf X}$ has been included to put in evidence
the conditioning variables used to evaluate GC.

On the other hand, an unnormalized version of it, i.e.
\begin{equation}\label{unnorm}
\delta^u_{\bf X} (B \to \alpha)=\epsilon \left(x_\alpha |{\bf
X}\setminus B\right)-\epsilon \left(x_\alpha |{\bf X}\right),
\end{equation}
can be easily  shown to satisfy the following interesting property:
if $\{X_\beta\}_{\beta\in B}$ are statistically independent and
their contributions in the model for $x_\alpha$ are additive, then
\begin{equation}\label{deltaset}
\delta^u_{\bf X} (B \to \alpha)=\sum_{\beta \in B} \delta^u_{\bf X}
(\beta \to \alpha).
\end{equation}
We remark that this property does not hold for the standard
definition of Granger causality neither for entropy-rooted
quantities \cite{beggs}, due to the presence of the logarithm.

\begin{figure}[thpb]
\centering
\includegraphics[width=8cm]{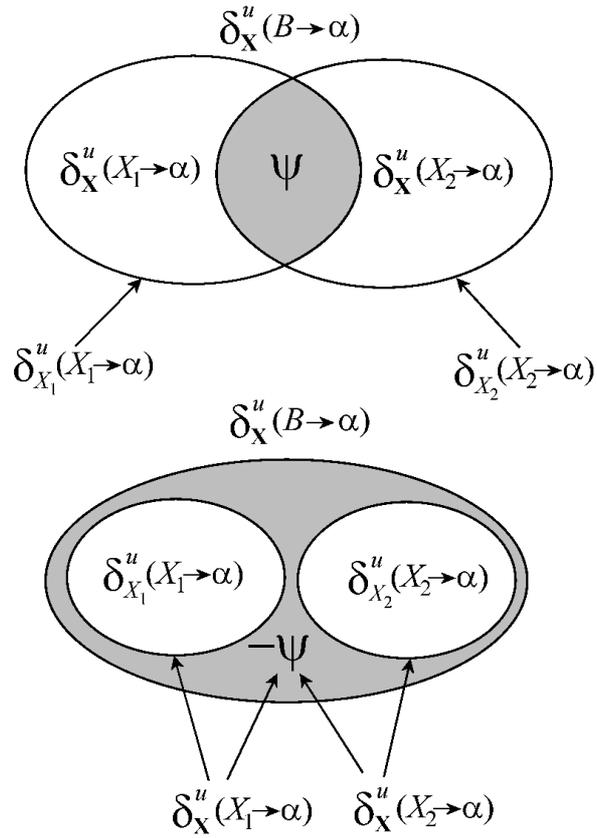}
      \caption{
Venn diagrams depicting the informational character
of two sources $B=\{ X_1 , X_2\}$ influencing a target $\alpha$.
Top: redundant source interaction ($\psi >0$); bottom:  synergetic
source interaction ($\psi  <0$). In both cases, the overall
unnormalized Granger causality from $B$ to $\alpha$   is decomposed
as: $\delta^u_{\bf X} (B \to \alpha)=\delta^u_{\bf X} (X_1 \to
\alpha)+ \delta^u_{\bf X} (X_2 \to \alpha) +\psi=\delta^u_{X_1} (X_1
\to \alpha)+ \delta^u_{X_2} (X_2 \to \alpha) -\psi$. }
      \label{venn}
\end{figure}

In order to identify the informational character of a set of
variables $B$, concerning the causal relationship $B \to \alpha$, we
remind that, in general, synergy occurs if $B$ contributes to
$\alpha$ with more information than the sum of all its variables,
whilst redundancy corresponds to situations with the same
information being shared by the variables in $B$ . We can render
quantitative these notions and define the variables in $B$ {\it
synergetic} if $\delta^u_{\bf X} (B \to \alpha) > \sum_{\beta \in B}
\delta^u_{{\bf X}\setminus B, \beta}(\beta \to \alpha)$, and {\it
redundant} if $\delta^u_{\bf X} (B \to \alpha) < \sum_{\beta \in B}
\delta^u_{{\bf X}\setminus B,\beta} (\beta \to \alpha)$. If GC is
computed conditioning on the whole set of time series ${\bf X}$, the
condition for synergy becomes $\delta^u_{\bf X} (B \to \alpha) <
\sum_{\beta \in B} \delta^u_{\bf X}(\beta \to \alpha)$ , and that
for redundancy becomes $\delta^u_{\bf X} (B \to \alpha) >
\sum_{\beta \in B} \delta^u_{\bf X}(\beta \to \alpha)$. All these
conditions are exemplified graphically in fig.(\ref{venn}) for the
simple case of two sources $B=\{X_1,X_2\}$. Note that the case of
independent variables (and additive contributions) does not fall in
the redundancy case neither in the synergetic case, due to
(\ref{deltaset}), as it should be.


Two analytically tractable cases are now reported as examples.
First, consider two stationary and Gaussian time series $x(t)$ and
$y(t)$ with $\langle x^2(t)\rangle=\langle y^2(t)\rangle=1$ and
$\langle x(t)y(t)\rangle={\cal C} $; they correspond, e.g., to the
asymptotic regime of the autoregressive system
\begin{eqnarray}
\begin{array}{ll}
x_{t+1}&=a x_t+b y_t +\sigma \xi^{(1)}_{t+1}\\
y_{t+1}&=b x_t+a y_t +\sigma \xi^{(2)}_{t+1},
\end{array}
\label{map}
\end{eqnarray}
where $\xi^{(i)}$ are i.i.d. unit variance Gaussian variables,
${\cal C}=2ab/(1-a^2-b^2)$ and $\sigma^2=1-a^2-b^2-2ab{\cal C}$.
Considering the time series
$z_{t+1}=A\left(x_t+y_t\right)+\sigma^\prime \xi^{(3)}_{t+1}$ with
$\sigma^\prime=\sqrt{1-2A^2(1+{\cal C})}$, we obtain for $m=1$:
\begin{equation}\label{z}
\delta^u_{\bf X} (\{x,y\} \to z)-\delta^u_{\bf X} (x \to
z)-\delta^u_{\bf X} (y \to z)=A^2({\cal C}+{\cal C}^2).
\end{equation}
Hence $x$ and $y$ are redundant (synergetic) for $z$ if ${\cal C}$
is positive (negative).

Let's then consider a nonlinear case, i.e. a target vector
given by $w_{t+1}=B\; x_t \cdot y_t+\sigma^{\prime \prime}
\xi^{(4)}_{t+1}$ with $\sigma^{\prime \prime}=\sqrt{1-B^2(1+2{\cal
C})^2}$, and using the polynomial kernel with $p=2$, we have
\begin{equation}\label{z}
\delta^u_{\bf X} (\{x,y\} \to z)-\delta^u_{\bf X} (x \to
z)-\delta^u_{\bf X} (y \to z)=B^2(4{\cal C}^2-1);
\end{equation}
$x$ and $y$ are synergetic (redundant) for $w$ if $|{\cal C}| < 0.5$
($|{\cal C}| > 0.5$).

The presence of redundant variables leads to under-estimation of
their Granger causality when the standard multivariate approach is
applied (as it is clear from the discussion above, this is not the
case for synergetic variables). Redundant variables should then be
grouped to get a reliable measure of Granger causality, and to
characterize interactions in a more compact way. As it is clear from
the discussion above, grouping redundant variables is connected to
maximization of the un-normalized Granger causality index
(\ref{unnorm}) and, in the general setting, can be made as follows.
For a given target $\alpha$, we call $B$ the set of the remaining
$n-1$ variables. The partition $\{A_\ell\}$ of $B$, maximizing the
total Granger causality
$$\Delta =\sum_\ell \delta^u_{\bf X} (A_\ell \to x_{\alpha}),$$
consists of groups of redundant variables.


As an example, we consider a redundant doublet of variables. Let
$h(t)$ be a zero mean and unit variance hidden Gaussian variable,
influencing two variables $x_i (t)=h(t-1)+ 0.5 \eta_i (t)$, $i=1,2$;
let $x_3=\eta_3 (t)$  and $w(t)=h(t-2) +0.1 \eta_0 (t)$ be another
variable which is influenced by $h$ but with a larger delay. The
$\{\eta\}$ variables are unit variance Gaussian noise terms. In
table 1 we report the value of the total Granger causality (to $w$)
for all the partitions of the three variables $x_1, x_2, x_3$.
\begin{table}[h]
\caption{Total Granger causality in the redundant example}
\label{table_example}
\begin{center}
\begin{tabular}{|c||c|}
\hline
partition& $\Delta$\\
\hline
 \hline
\{123\}& 0.88\\
\hline
\{12\}\{3\}& 0.88\\
\hline
\{13\}\{2\}& 0.18\\
\hline
\{23\}\{1\}& 0.18\\
\hline
\{1\}\{2\}\{3\}& 0.18\\
\hline
\end{tabular}
\end{center}
\end{table}
We note that in this example the correct partition, $\{12\}\{3\}$,
is the maximizer of the total Granger causality; however $\{123\}$
has the same value of $\Delta$. We note, however,
that merging variables is justified only if the total Granger
causality increases; therefore, in case of degeneracy, the partition
that must be chosen, among the maximizers, is the one with the
highest number of sets.
%
%

\subsection{Pairwise synergy index} The discussion in the previous
section suggests to quantitatively describe the informational
character of two variables $i$ and $j$, providing information for
the future state of the variable $x_\alpha$, by  the following
pairwise index:
\begin{eqnarray}
\begin{array}{lll}
\psi_\alpha (i,j)&=& \delta^u_{\bf
X\setminus j} (i \to \alpha)-\delta^u_{\bf X} (i \to \alpha)\nonumber\\
&=& \delta^u_{\bf X} (\{i,j\}\to \alpha)- \delta^u_{\bf X} (i\to
\alpha)-\delta^u_{\bf X} (j\to \alpha), \label{psi} \end{array}
\end{eqnarray}
where $\bf{X}$ is the set of conditioning variables. $\psi$ is
negative for increased unnormalized causality $i\to \alpha$ due to
the inclusion of $j$ in the conditioning variables (positive PSI
corresponds to redundancy), see figure (\ref{venn}) for a graphical
interpretation of $\psi$. Note that if i and j are statistically
independent and they cause $\alpha$ additively then PSI is zero,
differently from interaction information, where a common effect of
two causes induces a dependency among the causes that did not
formerly exist \cite{pearl88}.

\begin{figure}[thpb]
\centering
\includegraphics[width=8cm]{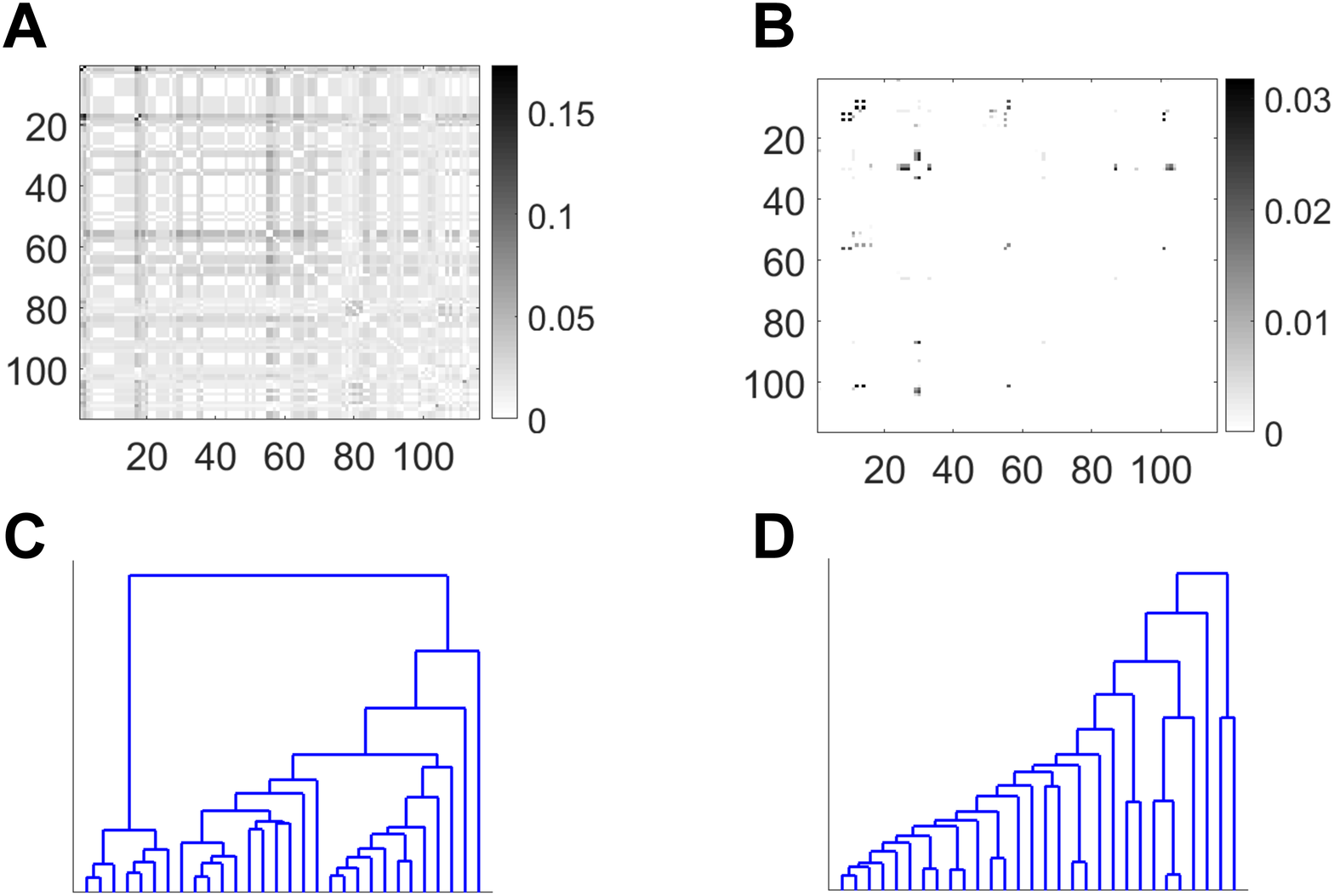}
      \caption{For the fMRI application, the matrices containing the values of redundant and synergetic duplets, and respective dendrograms, are depicted.
Top left (A): $\Psi_{r}$, redundant values for each pair; Bottom
left (C): dendrogram of the redundant matrix. Top right (B):
$\Psi_{s}$, synergetic values for each pair; Bottom right (D):
dendrogram of the synergetic matrix.}
      \label{fig3}
\end{figure}

Another interpretation of $\psi$ is given by the cumulant expansion
of the prediction error of $x_\alpha$:
\begin{equation}
\epsilon \left(x_\alpha |X_\alpha\right)-\epsilon \left(x_\alpha
|{\bf X}\right) =\sum_{B \subset {\bf X}} S(B). \label{cumulant}
\end{equation}
An important formula in combinatorics is the Moebius
inversion formula \cite{mat}, which allows to reconstruct $S(B)$
from equations (\ref{cumulant}) and (4). Calling $|n_B|$ and
$|n_\Gamma|$ the number of variables in the subsets $B$ and $\Gamma$
respectively,   and exploiting also the relation:
$$ \sum_{\Gamma \subset B} (-1)^{|n_\Gamma|}\; =0, $$
leads to the cumulant expansion:
\begin{equation}
S (B)=\sum_{\Gamma \subset B} (-1)^{|n_B|+|n_\Gamma|}\;\;
\delta^u_{B} (\Gamma \to \alpha).\label{cumulant1}
\end{equation}
The first order cumulant is then
\begin{equation}
S (i)=\delta^u_{i} (i \to \alpha),\label{cumulant2}
\end{equation}
the second cumulant is
\begin{equation}
S (i,j)=\delta^u_{ij} \left(\{ij\} \to \alpha\right)-\delta^u_{ij}
\left(i \to \alpha\right)-\delta^u_{ij} \left(j \to
\alpha\right),\label{cumulant3}
\end{equation}
the third cumulant is
\begin{eqnarray}
S (i,j,k)=\delta^u_{ijk} \left(\{ijk\} \to
\alpha\right)-\delta^u_{ijk} \left(\{ij\} \to \alpha\right)\nonumber
\\-\delta^u_{ijk} \left(\{jk\} \to \alpha\right)-\delta^u_{ijk} \left(\{ik\} \to
\alpha\right)\nonumber \\+\delta^u_{ijk} \left(i \to
\alpha\right)+\delta^u_{ijk} \left(j \to
\alpha\right)+\delta^u_{ijk} \left(k \to
\alpha\right),\label{cumulant4}
\end{eqnarray}
and so on. The index $\psi$ may then be seen as the order two
cumulant of the expansion of the prediction error of the target
variable; equation (\ref{cumulant1}) allows also the generalization
to higher order terms. Obviously $\psi$ also depends also on the
choice of the kernel, i.e. on the choice of the regression model.

\begin{figure}[thpb]
\centering
\includegraphics[width=8cm]{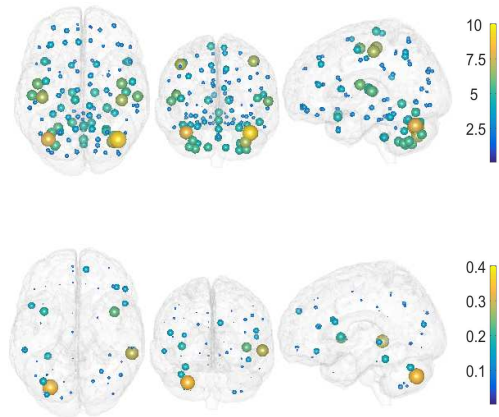}
      \caption{The strengths, sum of the redundant (top) and synergetic (bottom) contributions for each of the 116 brain regions, represented as spheres centered in the respective MNI coordinates. The size of the spheres is proportional to the depicted value.}
      \label{fig4}
\end{figure}

In order to go a step beyond, we observe that the pairwise synergy
index, as well as the cumulant expansion described above, is
dependent on the target $\alpha$. To get rid of this dependency, we
make the assumption that the essential features of the dynamics of
the system under consideration are captured using just a small
number of characteristic modes, and use principal components
analysis to obtain a compressed representation of the future state
of the system. Calling
$\{\xi_\lambda(t)\}_{\lambda=1,\ldots,n_\lambda}$ the time courses
of the largest $n_\lambda$ principal components of the whole system,
we define the pairwise synergy index:

\begin{equation}
\Psi(i,j)=\sum_{\lambda = 1}^{n_\lambda} \psi_\lambda
(i,j),\label{pca}
\end{equation}
obtained summing over the first $n_\lambda$ principal components
taken as targets.

The matrix $\Psi$ may be seen as a weighted network describing the
informational character of pairs of variables influencing the
future state of the system, zero entries meaning that the
corresponding variables provide independent information for the
future. The introduction of $\Psi$ allows to
describe redundancy and synergy in terms of  a weighted complex
network. In principle this makes complex
networks measures (e.g., modular decomposition) suitable to analyze these patterns in
large systems.

\section{Redundancy and synergy in resting state fMRI}

We now turn to investigate synergetic and redundant contributions in
large scale brain networks at rest. To this aim we consider resting
state functional magnetic resonance imaging (fMRI) recordings from
the Human Connectome Project (www.humanconnectome.org). Data are
acquired with a repetition time (TR) of 0.72 seconds, slice
thickness 2 mm, 72 slices, 2 mm isotropic voxels, 1200 frames. All
the parameters are reported in the project documentation. In this
study, the first 93 subjects in the 500 subjects release were used.
Data were preprocessed as described in \cite{prepro}. It is well
known \cite{seth} that the hemodynamic response function (HRF) can
confound the temporal precedence when lag-based methods are used to
infer directed connectivity in fMRI data, characterized by a slow
sampling rate. In order to address this issue in resting state data,
for which the onset of the HRF is not explicit, we used the blind
deconvolution approach described in \cite{wu2013}. The resulting
deconvolved BOLD signal was then averaged according to the 116
anatomical regions of the Automated Anatomical Labeling (AAL)
template \cite{aal}. Many choices of the parcellation are possible,
each one with its pros and cons, here we stick to this widely used
one.

The matrix $\Psi$, averaged over all the subjects, has been split in
two: the matrix of positive values, corresponding to redundancy and
defined as $\Psi_{r} = \theta \left( \Psi \right)$, $\theta$ being
the Heaviside function,  and the matrix of negative values,
corresponding to synergy and defined as $\Psi_{s} = \theta \left(-
\Psi \right)$. In figure (\ref{fig3}) we depict the two matrices and
the corresponding dendrograms. A very small number of pairs are
characterized by synergy, whilst the matrix of redundant pairs is
much more dense, and the corresponding dendrogram (differently from
the dendrogram of synergetic pairs) shows high values of modularity:
it follows that the modular decomposition of the redundancy matrix
(obtained considering the positive entries of $\Psi$) provides
correlated sets of variables acting as sources of information for
the dynamics of the system (modularity has been
evaluated, varying the resolution of the dendrogram, as described in
\cite{newman}).

In figure (\ref{fig4}) we depict the strength of the two matrices,
the synergy and the redundancy ones, on the brain. We see that the
pattern of redundancy is highly symmetric w.r.t. the two
hemispheres, whilst this is not the case for synergy, the highest
source of synergy being localized in the cerebellum.

\begin{figure}[thpb]
\centering
\includegraphics[width=8cm]{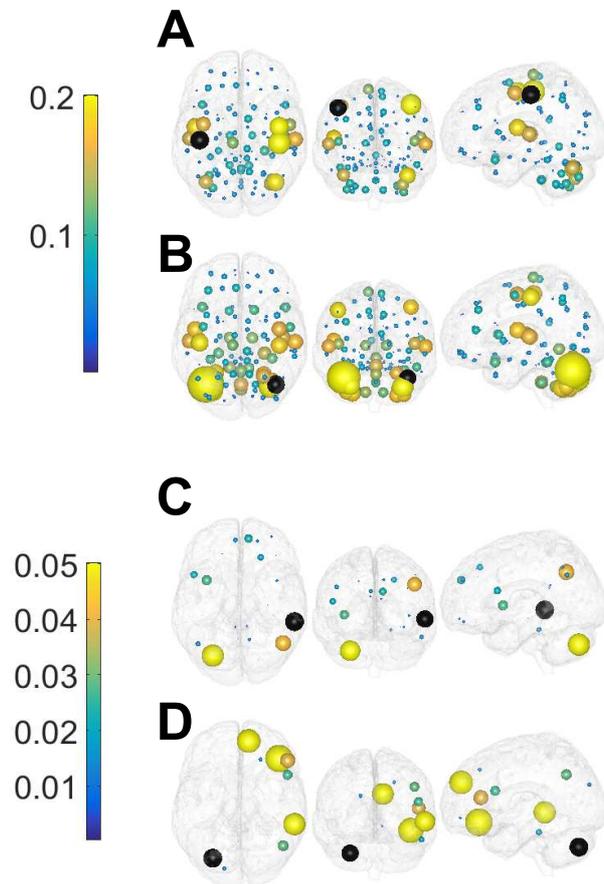}
      \caption{Regions which form redundant and synergetic duplets with representative regions, separately selected among cerebrum and cerebellar regions as those maximizing the overall informational contributions (see fig. \ref{fig4}). Regions are indicated by spheres centered in the MNI coordinates. The size of the spheres is proportional to the value of the duplet formed by the region and the representative one, depicted in black.
A: values of the redundant duplets involving the Left Postcentral
Gyrus B: values of the redundant duplets involving the Right Middle
Temporal Gyrus C: values of the synergetic duplets involving the
Left Postcentral Gyrus D: values of the synergetic duplets involving
the Left Cerebellar Crus II. All the
displayed values are significant as assessed using the closed form available for
Gaussian processes. Figures for all regions are available at \url{https://dx.doi.org/10.6084/m9.figshare.3101947}}
      \label{fig5}
\end{figure}

In figure (\ref{fig5}) we show some examples of redundant and
synergetic pairs of regions. In panel A we consider the cerebrum
area with highest redundancy strength, and plot the corresponding
absolute value of $\Psi$ w.r.t. all the other regions, showing that
homologous regions in the two hemispheres are redundant; this is
confirmed in panel B where a cerebellar region is considered. We
remark that the AAL partition allows to have homologous areas, thus
putting in evidence synergy-redundancy differences in the brain
localization. Redundancy is thus governed by contiguity in space as
well as by interhemispheric symmetry. In panel C and D, instead, we
consider the synergy between two given regions and all the others,
and observe that, accordingly, synergetic pairs are not homologous,
and pairs of regions with highest synergy correspond to a cerebrum
region and a cerebellar one on the opposite hemisphere. Maps for
all the regions are reported in the supplementary material.
\section{Conclusions}
\label{vvv} In this paper we have considered the inference, from
time series data, of the information flow between subsystems of a
complex network, an important problem in medicine and biology. In
particular we have analyzed the effects that synergy and redundancy
induce on the Granger causal analysis of time series. On one side,
we have shown that the presence of redundancy and synergy degrades
the performance of GC methods; on the other side, we have introduced
a frame for data analysis based on unnormalized Granger causality,
i.e. the reduction of variance of the residuals of each target
variables when candidate driver variables are included in the
regression model. Maximizing the total unnormalized Granger
causality leads to groups of redundant variables. We have introduced
a novel pairwise index of synergy, which for each pair of variables
measures how much they interact to provide better predictions of the
future state of the system, assumed to be synthesized in terms of a
small number of principal components. Such index can be seen as the
second cumulant in the expansion of the prediction error of the
target variables, to be compared with the expansion of the transfer
entropy in \cite{preExpansion} which provides the interaction
information as the second cumulant. The advantages provided by the
present cumulant expansion are (i) conceptual problems found in the
Gaussian case \cite{barrett} are avoided, and (ii) the nonlinearity
of $\Psi$ can be easily controlled by varying the kernel in the
regression model. A disadvantage of unnormalized GC is the
occurrence that the connection with information theory is lost, but
the aim of the present approach is to identify redundant and
synergetic circuits rather than quantifying the information flow in
the system. Our approach can be applied to any multivariate time
series data, here we have shown the efficacy of the proposed network
approach to resting state fMRI data. Redundancy appears to be more
bilateral than synergy, whilst the modular decomposition of the
redundancy matrix leads to correlated components of the system under
study. We have shown that redundancy is connected with space
contiguity of regions and interhemispheric symmetry, whilst synergy
occurs mainly between non-homologous pairs of regions in opposite
hemispheres.

Summarizing, we have addressed a novel framework
to study interdependencies among subcomponents of complex systems
from time series, which aim at highlighting redundant and
synergetic interactions. Whilst many examples of redundant
interactions have been reported in the literature, less attention
has been spent so far to synergetic interactions between variables
whose joint state influence the future of the system. Using the
pairwise synergy matrix we have shown that redundancy and synergy
display characteristic patterns in healthy brains at rest; such
patterns provide information about the system which are
complementary to those provided by standard multivariate analysis
tools, such as correlations and multivariate Granger
causality. A further step will be the assessment of alterations of these
patterns in pathologies and different brain states.

\addtolength{\textheight}{-12cm}   





\end{document}